\documentclass{article}

\usepackage{PRIMEarxiv}

\usepackage[utf8]{inputenc} % allow utf-8 input
\usepackage[T1]{fontenc}    % use 8-bit T1 fonts
\usepackage{hyperref}       % hyperlinks
\usepackage{url}            % simple URL typesetting
\usepackage{booktabs}       % professional-quality tables
\usepackage{amsfonts}       % blackboard math symbols
\usepackage{nicefrac}       % compact symbols for 1/2, etc.
\usepackage{microtype}      % microtypography
\usepackage{lipsum}
\usepackage{fancyhdr}       % header
\usepackage{graphicx}       % graphics
\graphicspath{{media/}}     % organize your images and other figures under media/ folder
\usepackage{epstopdf}% To incorporate .eps illustrations using PDFLaTeX, etc.
\usepackage{subfigure}% Support for small, `sub' figures and tables

\usepackage{amsmath}
\usepackage{algorithm}
\usepackage{algpseudocode}

\usepackage{natbib}% Citation support using natbib.sty
\usepackage{booktabs}
\usepackage[table]{xcolor} % For cell coloring
\usepackage{graphicx} % Required for including images

\usepackage{graphicx} % Required for including images

\usepackage{tikz}
\usetikzlibrary{shapes.geometric, arrows.meta, positioning,, calc}

\title{ Stagewise Anomaly Detection for E-Transaxle Quality Monitoring Using Wavelet and STFT Features
}

\author{
  Mohammad N. Bisheh \\
  H. Milton Stewart School of Industrial \& Systems Engineering \\
  Georgia Institute of Technology \\
  Atlanta, USA\\
   \And
  Rajesh Gupta \\
  Ford Motor Company \\
  USA \\
  \AND
  Qian Wang \\
  H. Milton Stewart School of Industrial \& Systems Engineering \\
  Georgia Institute of Technology \\
  Atlanta, USA\\
  \And
  Mohammad Babakmehr \\
  Ford Motor Company \\
  USA \\
  Current affiliation: Amazon Web Services\\
  \And
  Colin Brady \\
  Ford Motor Company \\
  USA \\
  \AND
  Parinaz Farajiparvar \\
  Ford Motor Company \\
  USA \\
  \And
  Saurabh Singh \\
  Ford Motor Company \\
  USA \\
  \And
  Kamran Payanabar \\
  H. Milton Stewart School of Industrial \& Systems Engineering \\
  Georgia Institute of Technology \\
  Atlanta, USA\\
}

\begin{document}
\maketitle

\begin{abstract}
This paper presents two interpretable machine-learning frameworks for quality screening of e-transaxle assemblies in electric vehicles: a Stagewise Wavelet Isolation Forest (SWIF) framework and a short-time Fourier transform (STFT)-based diagnostic framework. High-dimensional vibration signals acquired from front and back accelerometers are analyzed across multiple operating stages to capture stage-dependent vibration behavior. In the SWIF framework, signals are decomposed using a five-level Daubechies-4 discrete wavelet transform, and blockwise mean-squared coefficients are extracted from the selected wavelet detail level to obtain compact multiscale features. In the STFT-based framework, dominant-frequency trends are extracted from time-frequency representations and summarized through regression coefficients with respect to instantaneous motor speed. Anomaly detection models are trained using accepted production units under the assumption that only a small fraction of accepted assemblies contain latent defects, and their performance is evaluated using road-tested units with validated quality outcomes. Experiments on production and road-tested e-transaxle units show that both approaches provide interpretable diagnostic information, while SWIF achieves the most favorable balance between defect detection and false-positive control. Compared with the STFT-based method and alternative anomaly detectors, SWIF combined with Isolation Forest yields lower anomaly rates within the Accept population while identifying high-risk units from the Reject population. The stagewise structure further localizes anomalous behavior to specific operating conditions, supporting root-cause analysis and targeted process improvement. These results demonstrate that multiscale wavelet features, when coupled with one-class anomaly detection, provide a computationally efficient and practical tool for real-time e-transaxle screening, reducing unnecessary retesting while improving manufacturing efficiency and drivetrain reliability.

\end{abstract}

% keywords can be removed

\section{Motivation}
Electric vehicles (EVs) are at the forefront of sustainable transportation, offering substantial environmental benefits over traditional gasoline vehicles. By operating on electric power, EVs significantly reduce CO$_2$ emissions and other pollutants, thereby contributing to improved urban air quality and alleviating climate change \cite{hawkins2013comparative}. In addition, their higher energy efficiency, where electric motors convert over 60\% of electrical energy from the grid to power the wheels compared to about 20\% for the best internal combustion engines, results directly in lower energy consumption per mile traveled \cite{zhang2020noise, zhao2020design}. These advantages not only support environmental goals, but also reduce the overall resource consumption associated with transportation.

A key innovation within the EV drivetrain is the integration of the e-transaxle, which combines the electric motor and transmission mechanism into a single compact unit. This design offers significant advantages over traditional powertrains by distributing weight more evenly, lowering the vehicle's center of gravity, and reducing the number of moving parts, all of which enhance vehicle stability, handling, and overall safety \cite{ehsani2018modern}. Moreover, the precise control over torque provided by e-transaxles not only improves acceleration response but also plays a critical role in ensuring consistent vehicle performance. Reliable e-transaxle operation is essential, as it directly influences vehicle safety, drive quality, and customer satisfaction, while also enhancing manufacturing efficiency by reducing defects early in the production cycle \cite{qin2020noise}.

In contrast, traditional transmissions, typically mechanical in nature and paired with internal combustion engines, rely on conventional monitoring methods such as manual inspections, diagnostic tool readings, and occasional on-board diagnostics that focus primarily on transmission fluid characteristics (e.g., temperature, pressure, and contamination) \cite{yan2019weighted, liu2015failure}. These methods are further limited by their dependence on visual checks and basic vibration analyses, which are subject to human interpretation and lack real-time insight. Consequently, such approaches struggle to detect subtle anomalies in the early stages and often lead to inconsistent outcomes, highlighting the need for more advanced diagnostic methods.

\begin{figure*}[ht]
    \centering
    \includegraphics[width=0.5\linewidth]{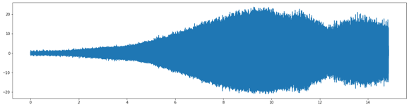}
    \caption{A sample vibration signal from E-transaxle}
    \label{fig:samplesignal}
\end{figure*}

The vibration data gathered across various operational phases of an e-transaxle, as depicted in Figures~\ref{fig:samplesignal}, offer detailed insights into drivetrain health and efficiency. For instance, the Speed Ramp phase illustrates a gradual rise in vibration amplitude as the vehicle accelerates, which can reveal issues such as bearing wear or motor misalignment before they result in failures. Likewise, marked shifts and higher-amplitude vibrations observed during Deceleration may point to regenerative braking effects and their influence on the drivetrain. Such granularity in vibration profiles is often absent or only partially available in traditional transmission systems, which tend to focus on obvious faults or significant drops in performance.

\begin{figure}[ht]
    \centering
    \includegraphics[width=0.5\linewidth]{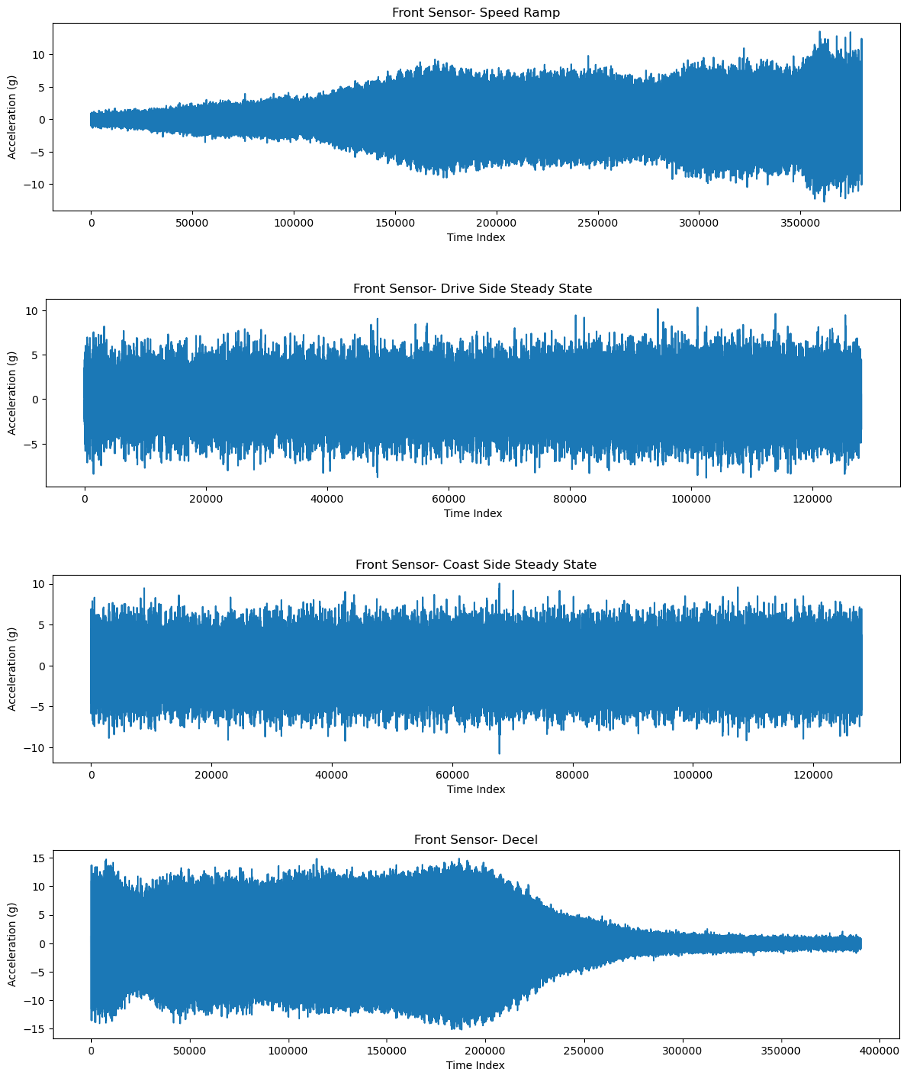}
    \caption{Comprehensive acceleration profiles of the E-transaxle recorded by the front sensor through four key operational phases: Speed Ramp, Drive Side Steady State, Coast Side Steady State, and Deceleration, demonstrating the dynamic response of the system}
    \label{fig:OriginalFront}
\end{figure}

This wealth of detailed data revolutionizes how engineers approach vehicle design and maintenance. Continuous feedback enables iterative refinements not only for performance but also for durability and noise mitigation, allowing the e-transaxle to be tailored to specific vehicle requirements and driving patterns.

Monitoring vibration signals effectively is crucial in many technical fields, especially in mechanical and automotive engineering, where the integrity and functionality of a system like an e-transaxle can be assessed based on these signals. The challenges associated with such signals typically include their high dimensionality, sparsity, and the presence of significant noise \cite{yan2018real,paynabar2016change} as it can be seen in Figure~\ref{fig:OriginalFront}. Transforming these signals from the time domain to the frequency domain can reveal the subtle details necessary for robust analysis and anomaly detection. There are various methods such as Order Analysis \cite{guo2022vibration}, Fourier Transform (FT) \cite{yang2020two}, Short Time Fourier Transform (STFT) \cite{wang2021automatic}, and Wavelet Transform \cite{choudhary2023fault} can be used to extract most informative features from both time and frequency domain that could be used to detect anomalies in production systems \cite{lang2021artificial}. The FT is a foundational method for signal processing that converts a time-domain signal into its constituent frequencies. This transformation is beneficial for analyzing periodic signals but can be limited when dealing with non-stationary signals where the frequency components vary over time \cite{yan2018real}. The STFT extends the capabilities of the Fourier Transform by adding a time dimension analysis. It involves dividing the signal into shorter stages and applying the Fourier Transform individually to each stage. This approach allows for observing how frequencies evolve over time, making it suitable for signals that change characteristics, such as during acceleration or deceleration phases in an e-transaxle. However, for signals characterized by high dimensionality and sparsity, the wavelet transform is particularly advantageous. Unlike the FT, which uses sinusoids, the wavelet transform employs wavelets with localized time and frequency characteristics. This method is powerful for analyzing transient phenomena and capturing details in signals with sudden changes, which is ideal for diagnostic purposes where the signals are sparse and embedded in noise \cite{shi2023process}.

Wavelet transformation stands out as a particularly powerful tool for analyzing signals that exhibit non-stationary behavior or contain high-frequency components transiently, which is often the case in mechanical systems like e-transaxles \cite{ni2022improved,xu2023mode}. Unlike the FT, which breaks down a signal into sine and cosine components, wavelet transformation uses wavelets, short, oscillating waveforms with limited duration. These wavelets are particularly effective in picking up anomalies, patterns, and changes in non-stationary signals because they vary in scale and position, adapting to the signal's local characteristics\cite{xu2023mode}. This method allows for the analysis of different frequency components with a resolution that matches their scales. The main advantage of using wavelet transformation in the context of vibration analysis is its ability to detect abrupt changes in the signal, such as mechanical faults or operational irregularities, which are often indicative of underlying issues. 

\begin{figure}[ht]
    \centering
    \includegraphics[width=0.95\linewidth]{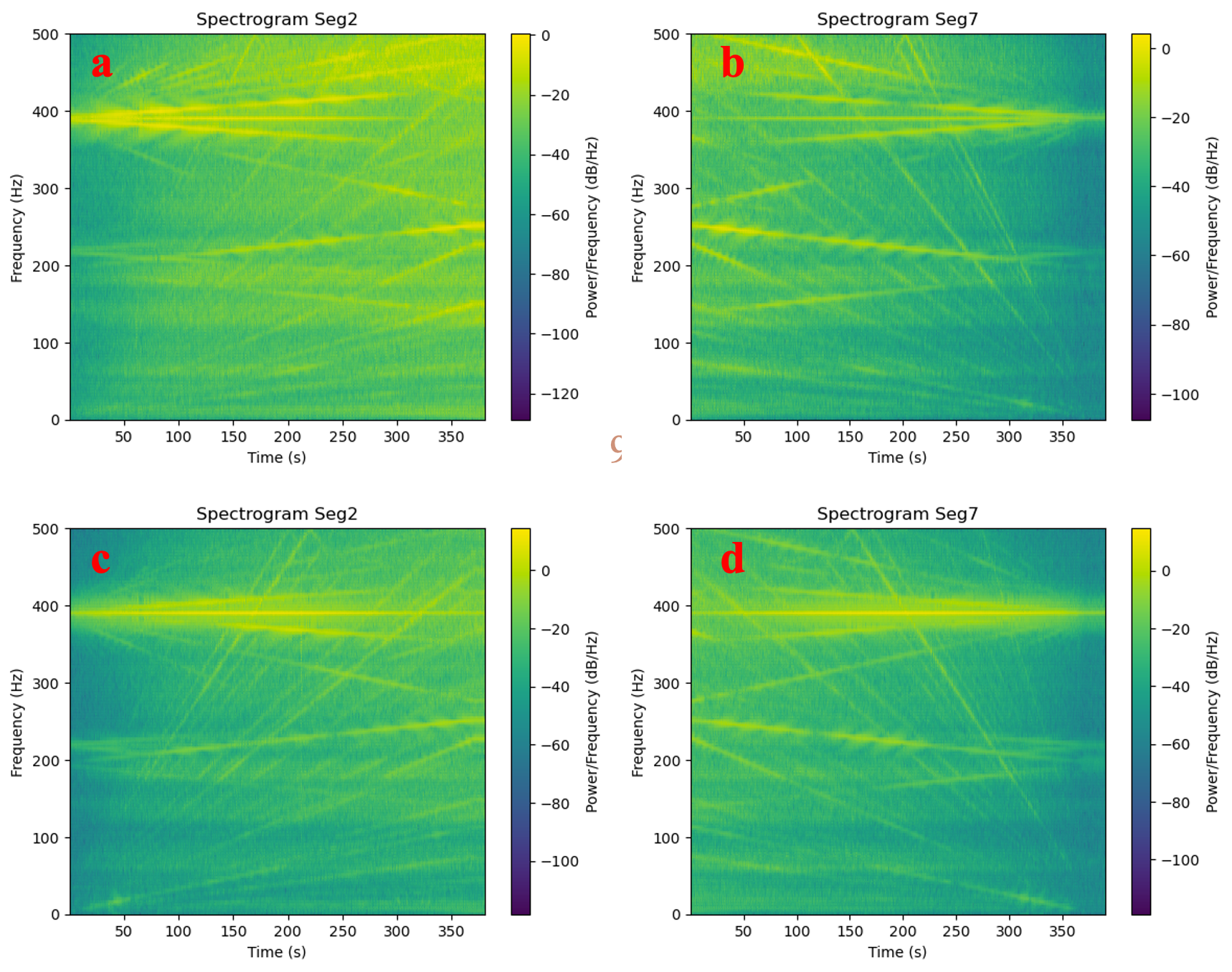}
    \caption{Spectrogram Analysis of Good and Defective E-Transaxles. This figure illustrates the vibration patterns in two operational stages (stage 1 and stage 4) of e-transaxles. Images (a) and (b) represent the vibration frequency and time distribution for a normal, functioning e-transaxle, while images (c) and (d) depict these characteristics for a defective unit. The spectrograms clearly visualize the differences in vibrational behavior, highlighting variations in frequency and timing that signal potential defects.}
    \label{fig:Wavelet scalogram}
\end{figure}

For instance, Figure~\ref{fig:Wavelet scalogram} vividly illustrates the contrast between a healthy and a defective e-transaxle. Each image presents a scalogram, a time-frequency map that tracks how vibration frequencies evolve over the course of operation. Parts (a) and (b) correspond to a normal, fully functional e-transaxle during two different stages, labeled here as Stage~1 and Stage~4. In these views, the color patterns remain relatively uniform and well-defined, reflecting stable vibration behavior. By contrast, parts (c) and (d) capture the same stages but in an e-transaxle that exhibits clear defects. Notice how the color spreads and clusters in more erratic patterns, signaling irregularities in both frequency and timing. This difference underscores how a defective unit deviates markedly from the expected vibration profile, thereby highlighting the root cause of the performance issue.

Employing wavelet transformation in our method for monitoring e-transaxle performance offers several benefits. Firstly, it enhances the sensitivity of the monitoring system to early signs of degradation or failure, allowing for timely maintenance and repair that can significantly extend the life of the equipment. Secondly, the detailed resolution provided by wavelet analysis helps in accurately pointing the type and location of faults within the system, which simplifies troubleshooting and reduces repair times. Lastly, because wavelet analysis can be applied in real-time, it supports advanced predictive maintenance strategies that optimize operational efficiency and reduce unexpected failures. By integrating wavelet transformation into our monitoring method, we provide a robust tool that improves reliability, enhances safety, and minimizes operational costs of systems relying on E-transaxles.

After extracting useful features from of the vibration signals of e-transaxles from both time and frequency domain, the focus now shifts towards leveraging these refined inputs for anomaly detection. Recent advances in unsupervised anomaly detection have also leveraged deep generative models, including diffusion-based approaches, to identify and localize anomalous patterns without requiring labeled defective samples \citep{moradi2025single,moradi2025rddpm}. This transition is a key moment where advanced data processing is combined with advanced analytical techniques. Machine learning (ML), with its robust capabilities in pattern recognition and statistical analysis, presents an ideal suite of methods to further this investigation \cite{kang2018machine}. By applying ML algorithms to the features isolated through wavelet transformation, it becomes possible to not only detect anomalies but also to predict potential failures before they appear. This integration is important for advancing e-transaxle diagnostics, enabling a proactive approach to maintenance that ensures reliability and efficiency in electric vehicle operations. These algorithms excel in handling HD data, a common characteristic of signal profiles in automotive engineering. This capability allows for scaling the anomaly detection process to handle data from numerous sensors simultaneously without a loss in performance. ML models, once trained, can evaluate data in a fraction of the time it takes human analysts or conventional statistical methods, leading to significant improvements in the efficiency of the diagnostic processes\cite{kang2018machine}. This rapid processing capability is vital for real-time monitoring systems, where quick detection of potential issues can prevent severe failures and reduce operational downtime.

iForest with linear time complexity is particularly well-suited for detecting anomalies in datasets where abnormal instances are few and different from the majority \cite{liu2008isolation}. Unlike other techniques that attempt to model the normal points in order to identify outliers, iForest isolates anomalies by randomly selecting a feature and then randomly selecting a split value between the maximum and minimum values of the selected feature \cite{xu2023deep}. This method is fundamentally more efficient in datasets with a large proportion of normal data and a small number of anomalies, as it requires fewer splits to isolate these anomalies, thereby speeding up the detection process and reducing the computational load \cite{liu2008isolation, xu2023deep}. iForest typically exhibits a lower false positive rate compared to methods like One Class SVM and Local Outlier Factor (LOF), especially in applications involving complex signal data. This precision derive from iForest’s unique isolation mechanism, which is less influenced by the underlying distribution of the data and more by the structural properties of the anomalies themselves. Consequently, it avoids the common risk of overfitting to the "noise" in the data, which can often lead to misclassification of normal instances as anomalies in other models.

Despite its effectiveness, iForest remains a relatively simple model in terms of implementation and parameter tuning, making it accessible for widespread use in industrial applications. This simplicity also extends to the interpretability of its results; the path lengths in the trees can give direct insights into the nature of the detected anomalies, which is invaluable for engineers seeking to understand the underlying causes of faults. This clarity is crucial for maintaining the trust of operators and engineers in automated monitoring systems and for facilitating further refinement of the diagnostic models based on operational feedback. These advantages make iForest a particularly robust choice for integrating with wavelet transformation techniques in the monitoring of e-transaxle performance, ensuring both high detection accuracy and operational efficiency in the anomaly detection processes.

\section{Problem description }

In this case study, we address the critical challenge of detecting Noise, Vibration, and Harshness (NVH) anomalies within EV drivetrains, with particular emphasis on e-transaxles integrated systems that combine the electric motor, gear reduction unit, and differential into a compact assembly. Given their central role in power transmission, e-transaxles are especially vulnerable to mechanical faults that present as subtle variations in vibration behavior \cite{nyman2025investigating,newcomb2023brief}. Even minor deviations, such as gear misalignment, bearing wear, or rotor imbalance, can generate early-stage NVH signatures that, if left unrecognized, may escalate into significant mechanical failures. Such latent faults not only compromise drivability and long-term durability, but can also result in expensive service interventions and erode customer satisfaction \cite{bleger2024automotive,newcomb2023brief}.

Despite the criticality of early anomaly detection, the automotive industry continues to rely heavily on standardized End-of-Line (EOL) inspection procedures \cite{liang2025transfer}. These inspections, while effective for capturing overt defects, often lack the sensitivity to identify more subtle or transient fault signatures, particularly in systems as compact and dynamically complex as e-transaxles. Conventional EOL protocols typically involve manual assessments or basic vibration analyzes based on scalar thresholds or summary statistics \cite{colledani2018cyber, moon2019noise,su2024correlation}. These approaches, however, are inadequate to capture the subtle multiscale nature of evolving NVH signals, and their effectiveness is further undermined by operator subjectivity and measurement variability.

As EV architectures evolve and the number of vehicle variants increases, the operational states of the drivetrains are becoming more diverse and their dynamic responses more variable \cite{colledani2018cyber}. This complexity, coupled with increasing production volumes and strict quality expectations, underscores the urgent need for more advanced data-driven diagnostic frameworks. In particular, there is a growing demand for automated fault detection methods capable of identifying early-stage NVH anomalies with high precision, minimal false positives, and operational scalability \cite{prakash2022bayesian}. A solution that satisfies these criteria would not only mitigate costly field failures and warranty claims but also enhance throughput efficiency and support more robust quality assurance practices in high-volume EV manufacturing environments.

Modern EOL testing for e-transaxles typically depends on a single inspection station that assigns a binary pass/fail label based on predefined vibration thresholds. However, this approach provides only a coarse assessment of the system’s dynamic behavior, as it often relies on fundamental vibration checks, such as time domain statistics or basic frequency domain analyses, that lack the sensitivity to capture subtle or transient fault signatures \cite{wheat2025correcting,song2022early}. These conventional techniques fail to deliver high resolution, real time insights into the complex and rapidly evolving vibration patterns inherent to e-transaxles, particularly under varying load conditions and operational regimes. As a result, accurately identifying the source of an anomaly is extremely difficult because the available data does not provide sufficient detail to distinguish between faults in gear interactions, bearing assemblies, or electromagnetic coupling \cite{makienko2024blind, wheat2025correcting}.

Even when certain key harmonic components (or 'orders') are monitored using control charts \cite{scheffer2004practical}, the static nature of thresholding renders the system inflexible to contextual changes, such as variations in production parameters or driving conditions. This rigidity often leads to two undesirable outcomes: borderline defects that escape detection and benign fluctuations that trigger false alarms \cite{prakash2022bayesian}. Such inconsistencies highlight the limitations of current EOL protocols and underscore the need for diagnostic frameworks that are more adaptive, fine-grained, and capable of learning from data over time. While recent advances in deep learning have shown promise in identifying complex vibration patterns and key orders \cite{liang2025transfer}, the black-box nature of these models stemming from their deep, nonlinear architectures poses significant challenges for interpretability. This lack of transparency hampers engineers’ ability to trace diagnostic decisions back to physical phenomena, limiting their utility in quality assurance and root-cause analysis.

These limitations in current diagnostic practices are further amplified by increasing production demands and the growing complexity of modern e-transaxles. As electric vehicle platforms evolve and diversify, the variability in operating conditions and mechanical configurations introduces additional challenges for EOL inspection. This shifting landscape creates a pressing need for a more robust and automated testing framework, one that not only scales to high throughput environments but also adapts to diverse signal patterns arising from multiple operational states. To be viable in production, such a method must process large volumes of HD vibration data rapidly and accurately, while maintaining the ability to distinguish subtle fault related anomalies from normal process variability. Achieving high sensitivity and high specificity is essential to ensure that true defects are reliably detected without triggering excessive false alarms or rework.

Meeting the growing demand for more precise and adaptive diagnostics requires signal processing methods that can capture subtle and rapidly changing features in complex vibration data. In this context, wavelet based techniques offer a promising alternative to traditional diagnostic approaches, particularly for detecting transient or spatially localized faults. Unlike standard time domain or frequency domain methods, wavelet transforms provide superior localization in both time and frequency, allowing signals to be analyzed at multiple resolutions simultaneously \cite{choudhary2023fault, shi2023process}. This capability makes wavelets especially effective for identifying short duration, low amplitude anomalies that are often masked by broader signal behavior or missed entirely by static thresholds and basic control charts. By decomposing vibration signals into localized components across different frequency bands, wavelet analysis can reveal hidden patterns that signal the early onset of mechanical degradation. This fine grained spectral representation is particularly valuable for isolating early stage defects such as gear tooth wear, bearing fatigue, or rotor imbalance, which are often overlooked by conventional techniques until significant damage has occurred.

In addition to advanced signal processing, the integration of ML algorithms plays a critical role in automating fault detection and ensuring adaptability to changing operational conditions. Unsupervised learning methods such as iForest, which identify outliers through random partitioning of the feature space, are particularly well suited for this task. These algorithms offer a substantial advantage over static rule based systems by significantly reducing the rate of false positives and eliminating the need for manually defined thresholds. Furthermore, machine learning enables real time analysis of large volumes of sensor data, which minimizes human error and accelerates the decision making process. As production data accumulates, the models can be continuously refined using feedback from the manufacturing line, allowing the system to improve its ability to distinguish meaningful anomalies from normal process variation. This adaptive learning capability is essential for maintaining high diagnostic accuracy in complex and evolving production environments.

Beyond the immediate improvement in detection accuracy, a more advanced diagnostic process offers significant benefits for overall production efficiency. Reducing dependence on repeated and costly road tests leads to higher throughput and minimizes the risk of delays or bottlenecks in the manufacturing line. Early and reliable identification of defects enables engineers to implement targeted corrective actions and refine component designs, thereby decreasing the likelihood of late-stage scrappage or failures occurring in the field. Over time, these improvements contribute to lower warranty costs and fewer product recalls. Just as importantly, they help build consumer confidence in the reliability and quality of electric drivetrains, ultimately enhancing the competitive position of manufacturers in the growing electric vehicle market.

\section{Dataset and Experimental Setup}
\label{sec:data}

The experimental framework utilizes two accelerometers mounted on the front and rear of the e-transaxle to capture vibration signals across different operating stages. These measurements enable the identification of vibrational anomalies associated with underlying mechanical defects or performance inefficiencies. To facilitate stage-specific analysis, the full time-series data from each test are partitioned into four distinct stages, each corresponding to a specific operational condition of the e-transaxle. This segmentation results in eight vibration signals per test, defined by the combination of sensor location and stage (e.g., Front–S1, Back–S4), enabling a structured and localized analysis of vibration behavior under varying operating conditions. An example of all four stages captured by front accelerometer is shown in figure \ref{fig:OriginalFront}.

stages 1 and 4, corresponding to the speed ramp-up and deceleration phases respectively, are designed to evaluate the e-transaxle’s response to dynamic changes in load and speed. stage 1 involves the application of a positive torque, simulating the conditions experienced during vehicle acceleration, and is intended to assess the e-transaxle's capacity to handle increasing mechanical demand. In contrast, stage 4 applies a negative torque to emulate deceleration, thereby testing the system’s ability to dissipate energy and transition smoothly to lower operational states.

stages 2 and 3 are associated with constant speed operation and are critical for assessing the e-transaxle’s performance under steady load conditions. These stages simulate typical driving scenarios in which the vehicle maintains a consistent speed, allowing for the evaluation of system stability and its tolerance to sustained mechanical stress. This part of the experimental setup is essential for capturing both transient and steady state characteristics across different operational profiles.

Our dataset consists of four categories: Accept, Reject, Accept RT, and Reject RT. Each e-transaxle is initially evaluated at a standard end-of-line test station and classified as either Accept or Reject. However, these labels do not represent perfect ground truth. Some units classified as Accept may still contain latent defects that are only identified later through warranty returns or field performance issues, while some units classified as Reject may ultimately be found to be defect-free.

Although the exact misclassification rates are unknown, the manufacturer maintains stringent quality-control standards and therefore the Accept population is expected to contain only a very small proportion of defective units. In contrast, the Reject population is expected to contain a mixture of truly defective and non-defective units. Consequently, the Accept dataset is treated as a predominantly normal population, while the Reject dataset is used primarily for evaluating the ability of the proposed framework to identify potentially defective units. This setting also reflects an important practical challenge in manufacturing, where conservative screening policies may lead to elevated false-positive rates and unnecessary retesting of otherwise acceptable products.

A subset of units that narrowly fail or pass the initial test, those on the borderline of being rejected/accepted, are selected for a more strict road test. Road testing is an expensive and time-consuming process conducted by experts, and its results are considered definitive, serving as our gold standard. Following the road test, each unit is reclassified as either "Reject RT" or "Accept RT". Ultimately, our analysis includes data from four stages of 1 to 4 across 935 Accept, 127 Reject, 9 "Accept RT", and 5 "Reject RT" e-transaxles.

\section{Proposed methods}
To address these challenges, we propose two machine-learning-based diagnostic frameworks to enhance the current engineering testing process by improving anomaly detection while reducing false positives. The first framework integrates wavelet transformation with ML-based anomaly detection, whereas the second uses STFT-derived features followed by the same anomaly detection techniques. Both approaches are described in detail in the following subsections.

Because road testing is costly and time-consuming, and because the initial test station produces a relatively high rejection rate, the goal is to develop a data-driven screening method that can identify truly defective units while reducing unnecessary rejections. We use the Accept population, which is assumed to consist predominantly of non-defective units, to train the anomaly detection models. The trained models are then applied to both Accept and Reject populations to detect true anomalies, while the available road-tested units are used as the validated test set for evaluating detection performance.

\subsection{Stage-Wise Wavelet–Isolation Forest (SWIF)}
The first proposed method includes a time-frequency domain aware feature extraction explained in subsection \ref{sec:feature} and the unsupervised anomaly detection explained in subsection \ref{sec:anomaly}. Figure \ref{fig:Method} also provides an overview of the proposed approach. The threshold $\tau$ is implicitly determined by the contamination parameter of the unsupervised model.

\subsubsection{Calculation of blockwise energy levels using wavelet transformation}
\label{sec:feature}

As shown in Figure \ref{fig:Method}, for each accelerometer (i.e. front and back) the analysis begins with the acquisition of HD, noisy time series data from the e-transaxle during operational tests across four stages, denoted as stages 1 through 4. These stages correspond to different operational conditions, such as acceleration, constant-speed cruising, and deceleration. To effectively extract subtle and localized vibrational features from the e-transaxle signals, we employ discrete wavelet transformation (DWT) using the Daubechies 4 (DB4) wavelet basis. 

\tikzset{
  stage/.style = {rectangle,
                  draw=teal!70!black,
                  fill=teal!8,
                  very thick,
                  rounded corners=3pt,
                  minimum width=4cm,
                  minimum height=1.3cm,
                  text width=2.8cm,
                  align=center,
                  font=\small},
  link/.style  = {-{Latex[length=3mm,width=2mm]},
                  very thick, draw=teal!70!black}
}
% --------------------------------------------------------------

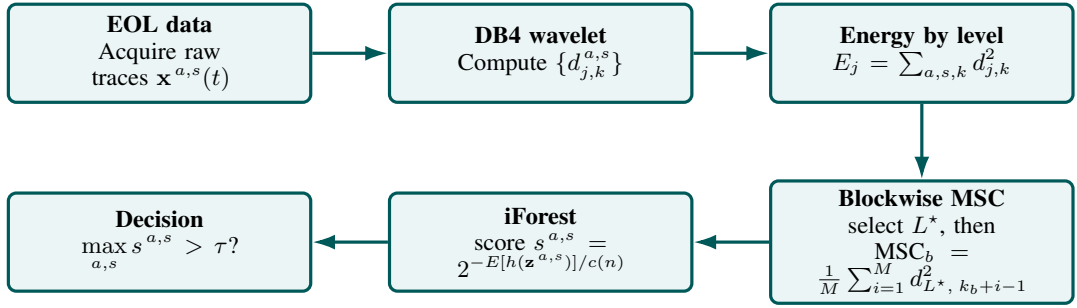
\begin{figure*}[ht]
\centering
\begin{tikzpicture}[node distance=1.0cm and 1.0cm]

% --- Stages ----------------------------------------------------
\node[stage] (raw)    {%
  \textbf{EOL data}\\
  Acquire raw traces $\mathbf x^{\,a,s}(t)$};

\node[stage, right=of raw] (dwt) {%
  \textbf{DB4 wavelet}\\
  Compute $\{d^{\,a,s}_{j,k}\}$};

\node[stage, right=of dwt] (energy) {%
  \textbf{Energy by level}\\
  $E_j=\sum_{a,s,k} d_{j,k}^{2}$};

\node[stage, below=of energy] (msc) {%
  \textbf{Blockwise MSC}\\
  select $L^\star$, then\\
  $\text{MSC}_b=\frac1M\sum_{i=1}^{M}d_{L^\star,\;k_b+i-1}^{2}$};

\node[stage, left=of msc] (score) {%
  \textbf{iForest}\\
  score $s^{\,a,s}=2^{-E[h(\mathbf z^{\,a,s})]/c(n)}$};

\node[stage, left=of score] (dec) {%
  \textbf{Decision}\\
  $\displaystyle \max_{a,s}s^{\,a,s}>\tau$?};

% --- Arrows ----------------------------------------------------
\draw[link] (raw)    -- (dwt);
\draw[link] (dwt)    -- (energy);
\draw[link] (energy) -- (msc);
\draw[link] (msc)    -- (score);
\draw[link] (score)  -- (dec);

\end{tikzpicture}
\caption{Stage-wise wavelet–Isolation-Forest (SWIF) workflow.  
Raw end-of-line (EOL) traces $\mathbf x^{\,a,s}(t)$ from four operating stages and two accelerometers undergo a 5-level DB4 decomposition.  
Per-level energy $E_{j}$ identifies the most discriminative scale $L^\star$.  
Blockwise mean-squared coefficients (MSC) form feature vectors $\mathbf z^{\,a,s}$, which are scored by stage-specific Isolation Forest models.  
An e-transaxle is rejected if any stage score exceeds the threshold~$\tau$.}
\label{fig:Method}
\end{figure*}

The choice of DB4 is motivated by its ability to effectively capture both frequency and time localization, which is essential for identifying subtle, transient anomalies in vibration signals \cite{chang2010statistical}. The DB4 wavelet, a compactly supported orthogonal wavelet, is particularly known for its low spectral leakage and minimal overlap in the frequency domain, making it well suited for high-resolution signal decomposition in noisy environments \cite{strang1996wavelets,mallat2002theory}.

The DWT begins with the definition of a scaling function $\phi(t)$, which satisfies the dilation equation:
\begin{equation}
\phi(t) = \sqrt{2} \sum_{k=0}^{N} p_k \phi(2t - k),
\end{equation}
where $p_k$ denotes the scaling (low-pass) filter coefficients. The corresponding wavelet function $\psi(t)$ is then defined as:
\begin{equation}
\psi(t) = \sqrt{2} \sum_{k=1}^{N} q_k \phi(2t - k),
\end{equation}
where $q_k$ represents the high-pass wavelet filter coefficients. For the DB4 wavelet, the filter coefficients are given by:
\begin{align}
q_0 &= \frac{1 + \sqrt{3}}{4\sqrt{2}}, & 
q_1 &= \frac{3 + \sqrt{3}}{4\sqrt{2}}, \notag \\
q_2 &= \frac{3 - \sqrt{3}}{4\sqrt{2}}, & 
q_3 &= \frac{1 - \sqrt{3}}{4\sqrt{2}}.
\end{align}

Using Mallat’s pyramid algorithm \cite{mallat2002theory}, the original signal $x(t)$ is decomposed across multiple scales:
\begin{equation}
x(t) = \sum_{m=1}^{L} \sum_{k=1}^{n} d_{mk} \psi_{mk}(t) + \sum_{k=1}^{n} a_{Lk} \phi_{Lk}(t),
\end{equation}
where $d_{mk}$ are the detail coefficients capturing high frequency content at scale $m$ and location $k$, and $a_{Lk}$ are the approximation coefficients capturing low-frequency content at scale $L$.

In this study, we apply discrete wavelet transformation (DWT) using the Daubechies 4 (DB4) wavelet to decompose HD vibration signals from the e-transaxle into multiple levels of resolution. Each level captures signal characteristics at a specific frequency band, offering a localized time–frequency representation. To enhance interpretability and focus the analysis on diagnostically relevant content, we systematically evaluate the decomposition levels to identify the most informative scale for anomaly detection.

The identification process involves calculating the mean squared value of the wavelet detail coefficients at each level. This value quantifies the energy content at that scale, thereby indicating the presence and concentration of significant signal variations. The decomposition level exhibiting the highest discriminative power across different operational conditions is retained for further analysis, while coefficients from other levels are discarded. The reconstructed signal from the selected detail coefficients enables us to isolate anomaly-relevant features while filtering out irrelevant noise. This process leverages the localization capability of the DB4 wavelet to maintain sensitivity to both frequency and temporal features within each operational stage.

To further reduce the dimensionality of the decomposed signals and to construct robust features for subsequent analysis, we stage the retained detail coefficients into fixed-length, non-overlapping windows and compute the mean squared value within each window. This step transforms thousands of time-domain observations into a compact set of statistically informative features that preserve sensitivity to localized anomalies.

Mathematically, the mean squared value at each decomposition level \( j \) is computed as follows:
\begin{equation}
\text{Mean Squared}_{j} = \frac{1}{N_j} \sum_{k=1}^{N_j} d_{jk}^2,
\end{equation}
where \( d_{jk} \) denotes the detail coefficient at level \( j \) and position \( k \), and \( N_j \) is the total number of coefficients at that level. This computation is performed for each decomposition level and for all e-transaxle units in the dataset.

To evaluate the discriminative power of each level, the computed mean squared values are visualized using box plots across four diagnostic categories: "Accept", "Reject", "Accept RT", and "Reject RT". These visualizations reveal which decomposition level provides the clearest separation between categories, allowing for the identification of the most informative level for anomaly detection. By selecting the level with the greatest discriminative capacity, we retain diagnostically relevant information while reducing dimensionality and noise.

This wavelet-based approach, combining DB4 decomposition with energy-based feature extraction, provides a compact and interpretable representation of the vibration signals. It serves as the foundation for the downstream anomaly detection framework, enabling early identification of subtle defects in a computationally efficient and scalable manner.

After identifying the most informative decomposition level through the wavelet-based energy analysis, the next step involves feature engineering to facilitate effective anomaly detection. This stage is designed to reduce data dimensionality while preserving the critical characteristics necessary for distinguishing between normal and faulty operational behavior. Moreover, it aims to enhance the interpretability of the diagnostic results, thereby enabling more precise root cause analysis.

The feature extraction strategy consists of partitioning the detail coefficients \( d_{jk} \) at the selected decomposition level \( j \) into non-overlapping blocks of fixed length and computing the mean squared value within each block. Specifically, given a total of \( N_j \) detail coefficients at level \( j \), the signal is divided into \( B \) blocks, each containing \( M \) coefficients. The mean squared value for the \( b \)-th block is computed as:
\begin{equation}
\text{Mean Squared}_{b} = \frac{1}{M} \sum_{i=1}^{M} d_{j,\,k_b+i-1}^2,
\end{equation}
where \( k_b = (b - 1) \times M + 1 \), \( b = 1, 2, \ldots, B \), and \( M \) is the block size. For example, when \( M = 1000 \) and \( N_j = 24,\!700 \), this process yields \( B = 25 \) summary statistics that compactly represent the localized energy distribution of the vibration signal across the stage.

These blockwise mean squared features serve as compressed yet informative representations of the original time series data, capturing localized variations that may correspond to incipient faults. By converting the HD wavelet coefficients into a lower-dimensional set of statistically meaningful features, this method enables scalable and interpretable downstream analysis while preserving sensitivity to spatially localized anomalies within each stage of the e-transaxle signal.

The resulting mean squared values are employed as input features for the anomaly detection framework. Each feature encapsulates localized energy from a specific block of the wavelet-transformed signal and reflects critical aspects of the underlying vibration behavior. Collectively, these features provide a compact yet expressive representation of the signal, enabling the detection of subtle deviations from normal operational patterns with high precision.

This dimensionality reduction not only enhances computational efficiency and facilitates the application of machine learning algorithms, but also focuses the analysis on the most informative regions of the signal. The blockwise calculation is applied across all stages of the selected decomposition level, offering a structured and stage specific view of the e-transaxle’s condition. Each mean squared feature corresponds to a localized temporal window within a stage and can be traced back to particular operational states or mechanical subsystems. This mapping enhances interpretability and supports targeted diagnostics.

This structured approach not only supports early detection of potential failures but also improves the ability to diagnose their root causes. By correlating abnormal features with specific operational behaviors or mechanical conditions, the methodology provides actionable insights into which components or dynamic elements of the e-transaxle may be malfunctioning. Consequently, the technique not only streamlines the diagnostic process but also reinforces its interpretability, contributing to more effective design iterations and process improvements.

In the anomaly detection phase of the methodology, the features derived from the blockwise mean squared values are analyzed using the iForest algorithm. This unsupervised learning method isolates anomalies based on the idea that abnormal points are more susceptible to isolation through recursive partitioning. The algorithm is applied separately to each stage, allowing for a fine-grained assessment of the vibration signal under different operational conditions. Both front and back accelerometer data are considered independently across stages, yielding a robust and stage specific identification of anomalous behavior within the e-transaxle system.

\subsubsection{Anomaly Detection Using iForest}
\label{sec:anomaly}

Following the feature extraction process, in which the vibration signal is condensed into blockwise mean squared values from the most informative wavelet decomposition level, we apply the iForest algorithm for anomaly detection. This unsupervised ensemble based method is well suited for HD settings and is particularly effective in identifying data points that deviate from the majority by exploiting the principle of recursive partitioning.

iForest constructs multiple binary trees by randomly selecting a feature and a split value between its minimum and maximum. Anomalies, data points that are rare or exhibit extreme behavior, tend to be isolated more quickly (i.e. at shallower depths in the tree structure). For a given observation \( \mathbf{x} \), the anomaly score is defined as:
\begin{equation}
s(\mathbf{x}, n) = 2^{-\frac{E(h(\mathbf{x}))}{c(n)}},
\end{equation}
where \( E(h(\mathbf{x})) \) is the average path length from the root to the terminating node across all trees in the ensemble, \( n \) is the number of training samples used to build each tree, and \( c(n) \) is the expected path length of unsuccessful searches in a Binary Search Tree, approximated as:
\begin{equation}
c(n) = 2 H(n-1) - \frac{2(n-1)}{n},
\end{equation}
with \( H(i) \approx \ln(i) + \gamma \), where \( \gamma \) is the Euler-Mascheroni constant (\( \gamma \approx 0.5772 \)).

A score \( s(\mathbf{x}, n) \) close to 1 indicates a high likelihood of being an anomaly, while values near 0 suggest conformity with the majority distribution.

In our implementation, the iForest is independently trained and applied to each of the eight stages of vibration data, four from the front accelerometer and four from the back. The characteristic vector of each stage, made up of some mean squared values, is evaluated to produce an anomaly score. The decision rule for flagging an e-transaxle unit as anomalous is defined as:
\begin{equation}
\text{Anomaly Flag} = 
\begin{cases}
1, & \text{if any stage is identified as anomalous}, \\
0, & \text{otherwise}.
\end{cases}
\end{equation}

This stagewise strategy enhances localization of faults and ensures that even subtle stage specific anomalies are not masked by aggregate behavior. It also supports targeted diagnostics by identifying which operational stage(s) contributed to the anomaly classification.

This approach that is summarized in algorithm \ref{alg:SWIF} ensures that any e-transaxle showing potential issues in any stage is subjected to further inspection, thereby enhancing the reliability of the diagnostic process and preventing possible failures from being overlooked. This methodology section describes the rigorous process of anomaly detection applied to e-transaxle units, ensuring high accuracy and reliability in identifying units that might require further analysis or repair.

\begin{algorithm}[ht]
\caption{Stage-Wise Wavelet + Isolation-Forest (SWIF) Screening}\label{alg:SWIF}
\begin{algorithmic}[1]
\Require Vibration traces $\mathbf{x}^{\,a,s}(t)$, $a\in\{\text{front,back}\}$, $s=1,\dots,4$
\Ensure Flag $\mathcal{A}\in\{0,1\}$ (0 = normal, 1 = anomalous)

% \Statex \textbf{1.  Pre-processing}
\State Detrend and $z$-score $\mathbf{x}^{\,a,s}(t)$ for every $(a,s)$

% \Statex \textbf{2.  Wavelet feature extraction}
\State Apply a 5-level DB4 DWT $\;\rightarrow\;$ detail sets $\{d^{\,a,s}_{j,k}\}$
\State Select most discriminative level 
      $L^* = \arg\max_j \left( \frac{1}{N_j} \sum_{a,s} \sum_{k=1}^{N_j} d_{a,s,j,k}^2 \right)$
\State Partition $\{d^{\,a,s}_{L^\star,k}\}$ into blocks of $M$ points and store block energy  
      $\text{MSC}^{\,a,s}_b = \tfrac1M\sum_{i=1}^{M}d^{\,2}_{L^\star,k_b+i-1}$  
      \emph{(feature vector $\mathbf{z}^{\,a,s}$)}

% \Statex \textbf{3.  Anomaly scoring (per stage)}
\For{$(a,s)$}
   \State Train Isolation Forest $\mathcal F^{\,a,s}$ on healthy $\mathbf{z}^{\,a,s}$
   \State Compute anomaly score $
          s^{\,a,s}=2^{-E[h(\mathbf{z}^{\,a,s})]/c(n)}$
\EndFor

% \Statex \textbf{4.  Unit level decision}
\State $\mathcal{A}\gets\bigl[\max_{a,s} s^{\,a,s}>\tau\bigr]$

\Return $\mathcal{A}$
\end{algorithmic}
\end{algorithm}

%\subsection{Preprocessing}

\subsection{STFT-based anomaly detection}
The second proposed method includes a pipeline that combines STFT \cite{mallat2008wavelet}, spline interpolation, and ordinary least-squares (OLS) regression on dominant spectral components.

Let \(x^{a,s}(t)\) denote the vibration signal from accelerometer  
\(a \in \{\mathrm{front},\mathrm{back}\}\) during stage  
\(s \in \{1,\dots,4\}\).  
After smoothing with a moving average length \(m\), we compute the STFT

\[
Z^{a,s}(t_n,f_m)=\!\!\sum_{\tau}
x^{a,s}(\tau)\,w(\tau-t_n)\,e^{-j2\pi f_m\tau},
\]

where \(w(\cdot)\) is a Hamming window and \(\lvert Z^{a,s}\rvert^2\) forms the spectrogram.  
For each time frame \(t_n\), we extract the dominant frequency and its magnitude,

\[
\begin{aligned}
f_{\mathrm{dom}}^{a,s}(t_n)&=\arg\max_{f_m}\lvert Z^{a,s}(t_n,f_m)\rvert,\\
A_{\mathrm{dom}}^{a,s}(t_n)&=\max_{f_m}\lvert Z^{a,s}(t_n,f_m)\rvert .
\end{aligned}
\]

Because the encoder signal delivers the rotational speed  
\(\omega(t)\) (RPM) at asynchronous timestamps, we align
\(\{f_{\mathrm{dom}}^{a,s}(t_n)\}\) and \(\omega(t)\) on a common grid via cubic B-spline interpolation:  

\[
\tilde f_{\mathrm{dom}}^{a,s}(t)=\operatorname{Spline}\bigl\{(t_n,f_{\mathrm{dom}}^{a,s}(t_n))\bigr\}(t).
\]

We model the relationship between dominant frequency and speed with an OLS fit,

\[
\tilde f_{\mathrm{dom}}^{a,s}(t)=
\beta_0^{a,s}+\beta_1^{a,s}\,\omega(t)+\varepsilon(t),
\]

and similarly for the interpolated dominant magnitude of the spline
\(\tilde A_{\mathrm{dom}}^{a,s}(t)\).
For each regressand we retain the slope and intercept,

\[
\mathbf{r}^{a,s}=
\bigl[\,
\beta_{1,f}^{a,s},\;
\beta_{0,f}^{a,s},\;
\beta_{1,A}^{a,s},\;
\beta_{0,A}^{a,s}
\bigr]^{\!\top}
% \in\mathbb{R}^{4}
.
\]

These coefficients are the only features visualized in the
box-plot analysis. 

The extracted features using SWIF and STFT could be used for anomaly detection such as iForest as we propose in section \ref{sec:anomaly}. However, to compare performance of the IF, we used the LOF, and a one-class SVM as a benchmark.

% \subsection{Anomaly detection}

\section{Results and discussion}

As mentioned in \ref{fig:Method}, both SWIF and STFT methods are implemented over the real dataset explained in \ref{sec:data}. The goal is to detect true anomalies while reducing costs by reducing false positive rate. The result of each method are discussed in sections \ref{sec:SWIF Result} and \ref{sec:STFT Result}.

\subsection{Results for the SWIF Model}
\label{sec:SWIF Result}

\begin{figure*}[ht]
    \centering
    \includegraphics[width=0.78\linewidth]{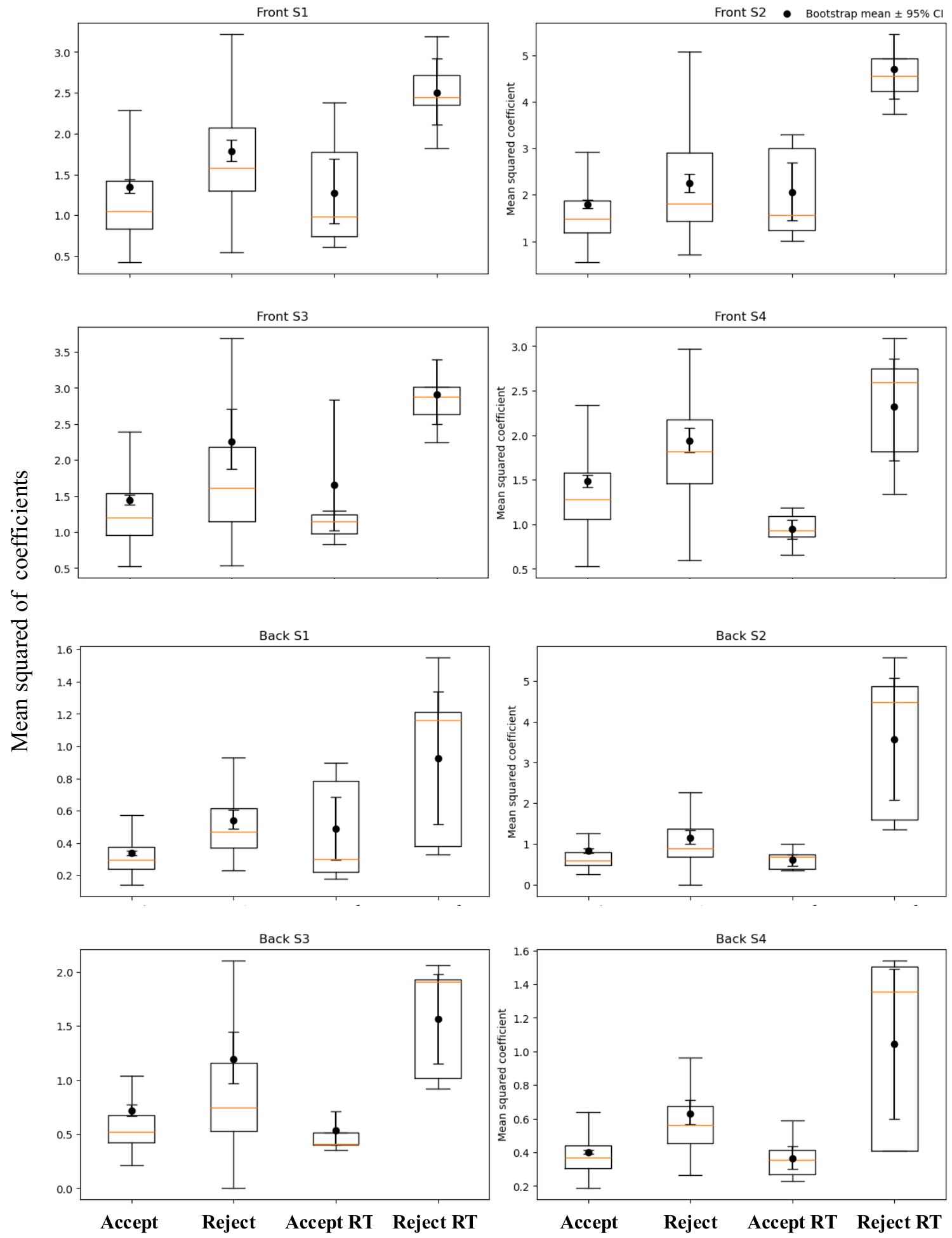}
    \caption{Box plots of mean squared wavelet coefficients across four stages for each of the front and back accelerometers. Vibration profiles for four categories Accept, Reject, Accept RT, and Reject RT highlighting variability and anomaly patterns under different test conditions.}
    \label{fig:WaveletFrontBack}
\end{figure*}

Figure~\ref{fig:WaveletFrontBack} presents box plots of the blockwise mean squared wavelet coefficients across eight stage–accelerometer pairs, comparing four quality categories of Accept, Reject, Accept RT, and Reject RT. These visualizations illustrate how the energy content of vibration signals, extracted via DB4 wavelet transformation, varies across operational conditions and quality states.

To quantify uncertainty, particularly in the small-sample road-test groups, two complementary approaches are employed. For the boxplot visualizations, bootstrap-based confidence intervals (CIs) are overlaid to characterize the variability of the estimated statistics. Specifically, for each group, resampling with replacement is performed and the statistic of interest, specifically the mean, is recomputed across a large number of bootstrap replicates. The resulting empirical distribution is used to construct percentile-based confidence intervals, which are shown as error bars. While boxplots summarize the spread and central tendency of the observed data, the bootstrap intervals provide an estimate of the stability of these statistics under sampling variability, enabling more reliable interpretation of differences across groups.

Across most front sensor stages, the Reject and Reject RT categories exhibit higher central tendencies and greater variability compared to the Accept and Accept RT categories. This trend is particularly pronounced in Front S2 and Front S4, where elevated values and wider spreads indicate increased vibration energy, consistent with mechanical defects such as imbalance or gear-mesh irregularities. In contrast, stages such as Back S1 and Back S3 show tighter distributions for Accept units, reflecting stable operation.

A clear distinction is also observed between front and back accelerometers. While both sensors capture elevated responses in defective units, the front accelerometer consistently records higher energy levels, suggesting stronger sensitivity to fault-induced vibrations due to structural or load-path differences within the e-transaxle.

Units in the Accept RT group exhibit distributions largely overlapping with Accept units, indicating borderline cases that pass road validation despite initial screening failures. Conversely, Reject RT units closely align with Reject behavior, reinforcing their classification as defective. The presence of extreme values, particularly in Front S4, suggests localized fault conditions that significantly amplify vibration energy. These observations confirm that the selected wavelet features effectively capture discriminative multiscale vibration characteristics associated with e-transaxle health states.

\begin{table*}[ht]
\centering
\small
\setlength{\tabcolsep}{5pt}
\renewcommand{\arraystretch}{1.2}
\caption{Classification performance on the road-tested dataset. Models were trained using only the Accept population with a contamination rate of 0.005 and evaluated using the validated Accept RT and Reject RT units.}
\label{tab:swif_rt_metrics}
\begin{tabular}{|c|c|c|c|c|c|c|c|c|c|}
\hline
\textbf{Detector} & \textbf{Total} & \textbf{TP} & \textbf{TN} & \textbf{FP} & \textbf{FN} & \textbf{Precision} & \textbf{Recall} & \textbf{F1-Score} & \textbf{Accuracy} \\
\hline
iForest & 14 & 3 & 9 & \textbf{0} & 2 & \textbf{1.000} & 0.600 & \textbf{0.750} & \textbf{0.857} \\
\hline
LOF & 14 & 3 & 6 & 3 & 2 & 0.500 & 0.600 & 0.545 & 0.643 \\
\hline
One-Class SVM & 14 & 5 & 5 & 4 & 0 & 0.556 & \textbf{1.000} & 0.714 & 0.714 \\
\hline
\end{tabular}
\end{table*}

\begin{table*}[ht]
\centering
\small
\setlength{\tabcolsep}{5pt}
\renewcommand{\arraystretch}{1.2}
\caption{Anomaly rates estimated for the production Accept and Reject populations. Models were trained using the Accept population under the assumption that fewer than 0.5\% of accepted units contain latent defects.}
\label{tab:swif_anomaly_rates}
\begin{tabular}{|c|c|c|c|c|}
\hline
\textbf{Detector} & \textbf{Group} & \textbf{Total} & \textbf{Flagged as Anomaly} & \textbf{Anomaly Rate} \\
\hline
iForest & Accept & 935 & 13 & \textbf{0.0139} \\
\hline
iForest & Reject & 127 & 13 & 0.1024 \\
\hline
LOF & Accept & 935 & 19 & 0.0203 \\
\hline
LOF & Reject & 127 & 12 & \textbf{0.0945} \\
\hline
One-Class SVM & Accept & 935 & 156 & 0.1668 \\
\hline
One-Class SVM & Reject & 127 & 78 & 0.6142 \\
\hline
\end{tabular}
\end{table*}

Table~\ref{tab:swif_rt_metrics} summarizes the classification performance of the three anomaly detection algorithms on the road-tested dataset, which serves as the most reliable source of ground-truth labels in this study. To reflect the manufacturing environment, all models were trained exclusively using the Accept population with a contamination rate of 0.005 for each individual sensor -- stage model, corresponding to the assumption that only a very small fraction of accepted units contain latent defects. Under this setting, iForest achieved the best overall performance, yielding a precision of 1.00, an F1-score of 0.75, and an accuracy of 85.7\%. The perfect precision indicates that every unit flagged by iForest in the road-tested dataset corresponded to a true defect. Although its recall was 0.60, resulting in two missed Reject RT units, the model maintained the strongest balance between sensitivity and specificity among the evaluated approaches. 

LOF exhibited substantially lower performance, achieving an F1-score of 0.545 and an accuracy of 64.3\%. While its recall matched that of iForest, the presence of three false positives reduced both precision and overall classification accuracy. In contrast, One-Class SVM achieved perfect recall by identifying all Reject RT units; however, this improvement came at the expense of a larger number of false positives, leading to a precision of only 0.556. These results indicate that One-Class SVM adopts a much more aggressive anomaly boundary, favoring sensitivity over specificity.

The practical implications of these differences become evident when examining the anomaly rates reported in Table~\ref{tab:swif_anomaly_rates}. Only 1.39\% of Accept units were flagged as anomalous by iForest, closely aligning with the assumption that the Accept population contains very few defective units. Note that the 1.39\% is the detected anomaly rate after aggregating across multiple sensor-stage models and each of the models has contamination rate equal to 0.005. More importantly, only 10.24\% of the Reject population was identified as anomalous. This finding suggests that a relatively small subset of rejected units exhibits vibration signatures that are strongly indicative of actual defects, providing a substantial opportunity to reduce unnecessary inspections, rework, and scrap costs. LOF produced a similar anomaly rate in the Reject population (9.45\%); however, its weaker performance on the validated road-tested dataset reduces confidence in its ability to correctly prioritize truly defective units.

One-Class SVM produced markedly different behavior, flagging 16.68\% of Accept units and 61.42\% of Reject units as anomalous. While this strategy minimizes the risk of overlooking defective units, it would substantially increase the number of components requiring secondary inspection and therefore increase operational costs. Consequently, although One-Class SVM demonstrates the highest sensitivity, iForest provides the most favorable trade-off between defect detection capability and inspection burden. These results suggest that the proposed SWIF framework, when combined with iForest, offers a practical solution for identifying high-risk units while significantly reducing the volume of components requiring costly follow-up evaluation.

\begin{table*}[ht]
\centering
\small
\setlength{\tabcolsep}{5pt}
\renewcommand{\arraystretch}{1.2}
\caption{Classification performance using the selected wavelet sensor--stage pair (Front S4). Models were trained using the Accept population with a contamination rate of 0.05 and evaluated using the validated Accept RT and Reject RT units.}
\label{tab:swif_selected_metrics}
\begin{tabular}{|c|c|c|c|c|c|c|c|c|c|}
\hline
\textbf{Detector} & \textbf{Total} & \textbf{TP} & \textbf{TN} & \textbf{FP} & \textbf{FN} & \textbf{Precision} & \textbf{Recall} & \textbf{F1-Score} & \textbf{Accuracy} \\
\hline
iForest & 14 & 5 & 9 & 0 & 0 & \textbf{1.000} & \textbf{1.000} & \textbf{1.000} & \textbf{1.000} \\
\hline
LOF & 14 & 1 & 8 & 1 & 4 & 0.500 & 0.200 & 0.286 & 0.643 \\
\hline
One-Class SVM & 14 & 3 & 9 & 0 & 2 & \textbf{1.000} & 0.600 & 0.750 & 0.857 \\
\hline
\end{tabular}
\end{table*}

\begin{table*}[ht]
\centering
\small
\setlength{\tabcolsep}{5pt}
\renewcommand{\arraystretch}{1.2}
\caption{Anomaly rates estimated for the production Accept and Reject populations using the selected wavelet sensor--stage pair (Front S4).}
\label{tab:swif_selected_anomaly_rates}
\begin{tabular}{|c|c|c|c|c|}
\hline
\textbf{Detector} & \textbf{Group} & \textbf{Total} & \textbf{Flagged as Anomaly} & \textbf{Anomaly Rate} \\
\hline
iForest & Accept & 935 & 47 & 0.0503 \\
\hline
iForest & Reject & 127 & 16 & 0.1260 \\
\hline
LOF & Accept & 935 & 41 & 0.0439 \\
\hline
LOF & Reject & 127 & 17 & 0.1339 \\
\hline
One-Class SVM & Accept & 935 & 57 & 0.0610 \\
\hline
One-Class SVM & Reject & 127 & 10 & 0.0787 \\
\hline
\end{tabular}
\end{table*}

To maintain a generalizable anomaly detection framework, the primary SWIF analysis utilized features extracted from all sensor--stage pairs rather than selecting a subset of stages. In practice, the most informative operational stages are typically unknown a priori, and restricting the model to specific stages may reduce its ability to detect previously unseen fault modes. Therefore, incorporating information from all stages provides a more comprehensive representation of e-transaxle behavior and improves the robustness of the screening process across different defect mechanisms.

However, the stage-level wavelet analysis indicated that certain sensor--stages pairs exhibit substantially stronger discriminative power than others. In particular, Front S4 showed the clearest separation between normal and defective units across the four sample categories. To investigate whether prior engineering knowledge could improve detection performance, an additional experiment was conducted using only the Front S4 feature. The results are summarized in Tables~\ref{tab:swif_selected_metrics} and~\ref{tab:swif_selected_anomaly_rates}. Under this targeted configuration, iForest achieved perfect performance on the available road-tested dataset, correctly classifying all Accept RT and Reject RT units, resulting in 100\% precision, 100\% recall, and 100\% accuracy. One-Class SVM maintained strong performance with an F1-score of 0.75, whereas LOF exhibited substantially lower sensitivity and failed to identify most Reject RT units.

The improved road-test performance demonstrates the value of incorporating process-specific knowledge into the anomaly detection framework. To achieve this improvement, however, the contamination rate was increased from 0.005 to 0.05, allowing the models to classify a larger fraction of observations as anomalous. Consequently, the anomaly rate within the Accept population increased from approximately 1.4\% in the general SWIF model to 5.0\% for iForest. Despite this increase, the anomaly rate in the Reject population rose only modestly, from approximately 10.2\% to 12.6\%. These results suggest that a relatively small increase in inspection burden can yield substantially improved defect detection performance when the most informative operating stage is known.

From a deployment perspective, the two modeling strategies serve different purposes. The all-stage SWIF model is more general and less dependent on prior knowledge, making it suitable for broad production screening and for detecting previously unknown fault mechanisms. In contrast, the stage-specific model leverages engineering insight obtained from historical data and targeted analysis to achieve superior classification performance for known fault modes. Therefore, a practical implementation may employ the all-stage model as an initial screening tool, followed by a stage-focused model for detailed evaluation of units associated with known failure mechanisms.

\subsection{Results for the STFT-Based Model}
\label{sec:STFT Result}

\begin{figure*}[ht]
    \centering
    \includegraphics[width=1.05\linewidth]{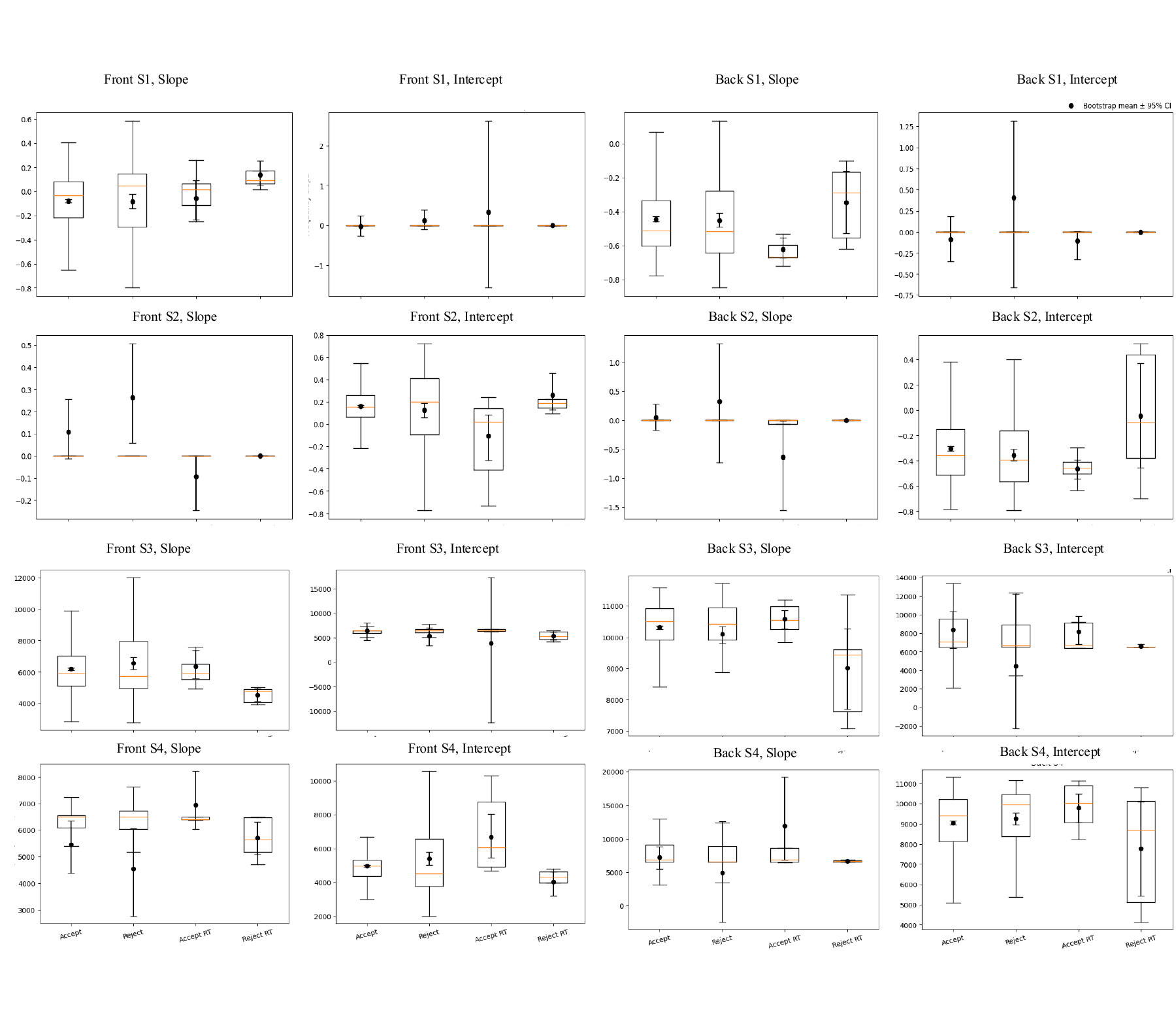}
    \caption{Box plots of STFT--RPM regression coefficients across eight stage-accelerometer pairs. Slopes and intercepts are compared for Accept, Reject, Accept~RT, and Reject~RT categories. Bootstrap mean and 95\% confidence intervals are overlaid to quantify uncertainty, particularly for small-sample road-test groups.}
    \label{fig:STFT_Box}
\end{figure*}

Figure~\ref{fig:STFT_Box} displays box plots of the regression coefficients of slope and intercept obtained by fitting the dominant frequency traces from STFT to the instantaneous RPM signal. Eight stage–accelerometer pairs are shown, comparing four quality categories of Accept, Reject, Accept~RT, and Reject~RT. The slope quantifies the rate at which the dominant frequency evolves with rotational speed, while the intercept represents the baseline frequency component. Together, these coefficients provide complementary descriptors of vibration behavior.

Across several stages, particularly Front~S2 and Back~S2, the Reject and Reject~RT categories exhibit broader distributions and higher variability in slope values compared to Accept units. This suggests that defective units experience more pronounced shifts in dominant frequency as speed increases, which is consistent with fault mechanisms such as gear-mesh irregularities or bearing clearance issues. However, the separation between Accept and Reject groups is less consistent across all stages compared to the wavelet-based features.

The intercept distributions show partially similar trends. In some stages, Reject and Reject~RT groups exhibit elevated baseline frequency values, indicating increased vibration energy even at lower rotational speeds. However, the overlap between categories remains substantial in several stages, indicating limited discriminative power of intercept features alone.

Although both front and back sensors capture similar qualitative patterns, their magnitudes and variability differ, reflecting differences in structural response and vibration transmission across sensor locations. Overall, the STFT-based coefficients capture global frequency trends but exhibit weaker separation between health states compared to wavelet-based features.

When all STFT-derived features from the eight sensor--stage pairs were included in the anomaly detection framework, the resulting models exhibited poor discrimination under realistic manufacturing assumptions. Similar to the SWIF framework, a contamination rate of 0.005 was initially adopted to reflect the expectation that only a very small fraction of accepted units contain latent defects. Under this setting, the STFT-based models failed to identify a sufficient number of defective units in the road-tested dataset. Increasing the contamination rate slightly improved recall; however, this came at the expense of a large increase in false-positive rates within both Accept and Reject populations. Such behavior is undesirable in a manufacturing environment because it would lead to excessive secondary inspections and unnecessary operational costs. Consequently, these results are not reported here.

However, inspection of the STFT boxplots revealed that several individual sensor--stage features exhibited noticeably stronger separation between quality categories than the remaining features. In particular, the Back S1 frequency-intercept feature demonstrated a consistent distinction between the Accept RT and Reject RT populations. Although some overlap remained, the Reject RT group generally exhibited elevated frequency-intercept values relative to the Accept RT group. Furthermore, the Reject production population displayed a similar trend to the Reject RT units, whereas the Accept population more closely resembled the Accept RT units. These observations suggest that the Back S1 frequency-intercept feature captures vibration characteristics associated with defective e-transaxles more effectively than most other STFT-derived features.

Table~\ref{tab:stft_rt_metrics} summarizes the anomaly detection performance obtained using only the Back S1 frequency-intercept feature. Among the evaluated methods, LOF achieved the best overall performance, producing a precision of 1.00, recall of 0.80, F1-score of 0.889, and classification accuracy of 92.9\%. The model correctly identified four of the five Reject RT units while generating no false positives within the Accept RT population. iForest produced slightly lower performance, achieving an F1-score of 0.75 and accuracy of 85.7\%. One-Class SVM successfully detected all Reject RT units but misclassified every Accept RT unit as anomalous, resulting in a very low precision of 0.357 and an overall accuracy of only 35.7\%.

The estimated anomaly rates for the production Accept and Reject populations are presented in Table~\ref{tab:stft_anomaly_rates}. LOF generated the lowest anomaly rate in the Accept population (7.59\%), while both LOF and iForest identified approximately 19.7\% of the Reject population as anomalous. Although these results demonstrate that carefully selected STFT features can provide useful diagnostic information, the overall performance remains less robust than that achieved by the SWIF framework. Unlike the wavelet-based approach, the STFT methodology relies heavily on identifying a small number of informative features and exhibits greater sensitivity to parameter selection. Therefore, while STFT-derived features may serve as supplementary indicators, the SWIF framework provides a more reliable and scalable solution for practical e-transaxle anomaly detection.

\begin{table*}[ht]
\centering
\small
\setlength{\tabcolsep}{5pt}
\renewcommand{\arraystretch}{1.2}
\caption{Classification performance of the STFT-based anomaly detection framework using only the Back S1 frequency-intercept feature. Models were trained using the Accept population and evaluated on the road-tested Accept RT and Reject RT units.}
\label{tab:stft_rt_metrics}
\begin{tabular}{|c|c|c|c|c|c|c|c|c|c|}
\hline
\textbf{Detector} & \textbf{Total} & \textbf{TP} & \textbf{TN} & \textbf{FP} & \textbf{FN} & \textbf{Precision} & \textbf{Recall} & \textbf{F1-Score} & \textbf{Accuracy} \\
\hline
iForest & 14 & 3 & 9 & 0 & 2 & 1.000 & 0.600 & 0.750 & 0.857 \\
\hline
LOF & 14 & 4 & 9 & 0 & 1 & \textbf{1.000} & \textbf{0.800} & \textbf{0.889} & \textbf{0.929} \\
\hline
One-Class SVM & 14 & 5 & 0 & 9 & 0 & 0.357 & 1.000 & 0.526 & 0.357 \\
\hline
\end{tabular}
\end{table*}

\begin{table*}[ht]
\centering
\small
\setlength{\tabcolsep}{5pt}
\renewcommand{\arraystretch}{1.2}
\caption{Estimated anomaly rates for the production Accept and Reject populations using the STFT-based framework with the Back S1 frequency-intercept feature.}
\label{tab:stft_anomaly_rates}
\begin{tabular}{|c|c|c|c|c|}
\hline
\textbf{Detector} & \textbf{Group} & \textbf{Total} & \textbf{Flagged as Anomaly} & \textbf{Anomaly Rate} \\
\hline
iForest & Accept & 935 & 94 & 0.1005 \\
\hline
iForest & Reject & 127 & 25 & 0.1969 \\
\hline
LOF & Accept & 935 & 71 & \textbf{0.0759} \\
\hline
LOF & Reject & 127 & 25 & 0.1969 \\
\hline
One-Class SVM & Accept & 935 & 344 & 0.3679 \\
\hline
One-Class SVM & Reject & 127 & 108 & 0.8504 \\
\hline
\end{tabular}
\end{table*}

\section{Conclusion}

This study demonstrates that integrating DB4 wavelet-based feature extraction with an Isolation Forest classifier yields an effective and computationally efficient framework for anomaly detection in e-transaxle assemblies. Comparative results show that the proposed SWIF workflow provides a more favorable balance between defect detection and false-positive control than both the STFT-based approach and alternative anomaly detection methods, including LOF and One-Class SVM. Under a realistic manufacturing assumption that only a small fraction of accepted units contain latent defects, the SWIF-iForest framework achieved perfect precision on the validated road-tested dataset while maintaining competitive recall and the highest overall accuracy among the evaluated methods. From a manufacturing perspective, the primary objective is not only to detect defective units but also to reduce unnecessary inspections and associated costs. The proposed framework addresses this challenge by identifying only a small subset of the Reject population as high-risk while maintaining a very low anomaly rate within the Accept population. These results suggest substantial opportunities to reduce secondary inspections, rework, and scrap costs without compromising product quality. In addition, reliable identification of defective units prevents faulty drivetrains from progressing to final vehicle assembly, thereby improving manufacturing robustness, product reliability, and customer satisfaction. Beyond classification performance, the proposed SWIF framework provides interpretable stage-level diagnostics through blockwise wavelet features, enabling localization of fault signatures across different operating conditions and facilitating targeted root-cause analysis. The superior performance of the wavelet-based approach also highlights the value of multiscale, time-localized signal representations for capturing physically meaningful vibration characteristics that are not adequately described by global frequency-trend descriptors derived from STFT. Consequently, SWIF offers both practical deployment benefits and actionable engineering insight for continuous quality improvement in high-volume manufacturing environments. Future work will extend the framework to additional e-transaxle architectures and operating conditions, investigate adaptive wavelet parameterization and automated feature selection strategies, and integrate real-time data streams for online monitoring and decision support. Such developments will further enhance manufacturing robustness and support the broader deployment of reliable electric-drive technologies.

\section*{Gen AI Use}
ChatGPT was used for grammar checking, text polishing, and code development. All ideas presented in this work were developed by the authors.

% \section*{Acknowledgments}
% This was was supported in part by......

%Bibliography
\bibliographystyle{unsrt}  
\bibliography{references}

\end{document}